# Observation of Optical and Electrical In-plane Anisotropy in High-mobility Few-layer ZrTe$_5$

Gang Qiu[1,4], Yuchen Du[1,4], Adam Charnas[1,4], Hong Zhou[1,4], Shengyu Jin[2,4], Zhe Luo[3,4], Dmitry Y. Zemlyanov[4], Xianfan Xu[3,4], Gary J. Cheng[2,4], Peide D. Ye[*,1,4]

[1] *School of Electrical and Computer Engineering, Purdue University, United States*

[2] *School of Industrial Engineering, Purdue University, United States*

[3] *School of Mechanical Engineering, Purdue University, United States*

[4] *Birck Nanotechnology Center, Purdue University, United States*

\* Corresponding author

E-mail: yep@purdue.edu.    Fax: 765-496-7443

# Abstract

Transition metal pentatelluride $ZrTe_5$ is a versatile material in condensed-matter physics and has been intensively studied since the 1980's. The most fascinating feature of $ZrTe_5$ is that it is a 3D Dirac semimetal which has linear energy dispersion in all three dimensions in momentum space. Structure-wise, $ZrTe_5$ is a layered material held together by weak interlayer van der Waals force. The combination of its unique band structure and 2D atomic structure provides a fertile ground for more potential exotic physical phenomena in $ZrTe_5$ related to 3D Dirac semimentals. However the physical properties of its few-layer form have yet to be thoroughly explored. Here we report strong optical and electrical in-plane anisotropy of mechanically exfoliated few-layer $ZrTe_5$. Raman spectroscopy shows significant intensity change with sample orientations, and the behavior of angle-resolved phonon modes at the Γ point is explained by theoretical calculation. DC conductance measurement indicates a 50% of difference along different in-plane directions. The diminishing of resistivity anomaly in few-layer samples indicates the evolution of band structure with reduced thickness. Low-temperature Hall experiment sheds lights on more intrinsic anisotropic electrical transport, with hole mobility of 3,000 and 1,500 $cm^2/V\cdot s$ along a-axis and c-axis respectively. Pronounced quantum oscillations in magneto-resistance are observed at low temperatures with highest electron mobility up to 44,000 $cm^2/V\cdot s$.

The discovery of graphene[1] began a new era of condensed-matter research because of its unique two-dimensional Dirac band structure, which hosts many profound physical phenomena such as the anomalous integer quantum Hall effect (IQHE) [2]. Since then great efforts have been made towards expanding the spectrum of topological materials and bringing many conceptual materials into reality. Transition metal pentatellurides such as $ZrTe_5$ and $HfTe_5$ have been widely studied in bulk form since early 1980's due to their anomalous resistivity peak and X-ray diffraction intensity peak at low temperature[3,4], large thermoelectric power[5], pressure-induced superconductivity[6,7], absence of a structural phase transition corresponding to resistivity anomaly[8], and chiral magnetic effect[9]. In recent years, $ZrTe_5$ research has been revived because of its non-trivial topological properties. Some theoretical predictions and experimental results[10,11] indicate that it is a 3D Dirac semimetal, a mimic of graphene with linear energy dispersion in all three directions. On the other hand, its monolayer form is also claimed to be a candidate of quantum spin Hall insulator[12,13], which is very rare among the natural compounds[14]. Shubnikov-de Haas oscillations[10,15,16], Zeeman Splitting[17,18], and fractional quantum Hall effect[19] were also observed in bulk $ZrTe_5$.

Meanwhile in recent years the 2D family has been expanded to a wide range of materials including narrow bandgap semiconductors (e.g. black phosphorus[20,21]), wide bandgap semiconductors (e.g. transition metal dichalcogenides[22]) and insulators (e.g. boron nitride[23]). Dirac semimetal $ZrTe_5$ is also a two-dimensional material, with neighboring atomics layer connected by van der Waals force. The interaction of Dirac fermion between interlayers may introduce more exotic physics in this material.

In this letter, we focus on in-plane anisotropic phenomena of few-layer ZrTe$_5$, which is a common property shared by a variety of other 2D materials. For example, black phosphorus has electrical[20] and thermal[24] conductivity anisotropic ratios of around 1.5 and 2, respectively. Black phosphorus also exhibits anisotropic infrared and Raman spectra[21]. Similar works were carried out on other 2D materials such as graphene[25] and TMDs[26–28]. However, a detailed investigation on anisotropic properties of few-layer ZrTe$_5$ is still absent. Here we report a systematic study on both optical and electrical anisotropy on few-layer ZrTe$_5$. The TEM image illustrates the quasi-one-dimensional structure of ZrTe$_5$, and the strong anisotropy originates from its unique atomic structure. Angle-resolved Raman spectra reveal anisotropic phonon dispersion for different Raman modes. This anisotropic Raman behavior is explained by group theory calculation[29]. By carefully designing device structures, we find that the DC conductance along the a-axis is 1.5 times larger than that along c-axis. Finally we carried out low temperature Hall experiment to measure carrier concentration and Hall mobility in different directions. Strong Shubnikov-de Haas oscillations were observed in high-mobility ZrTe$_5$ samples. This work not only gives an insightful understanding of the material anisotropic properties, but also provides a reliable means to determine the film orientations.

Fig 1(a) and (b) represent the side view and top view of ZrTe$_5$ lattice structure. One zirconium atom and three tellurium atoms are arranged into a tetrahedron. These primitives are repeated to form quasi-one-dimensional Zr chains that stretch along the crystalline a-axis. Zr atoms in adjacent chains are connected by two Te atoms in distorted angles in c-axis to form a 2D layer. The layers are then piled up in b-axis with a half-period shift between two neighboring layers. Figure 1 (c) shows the TEM image of (010) surface. The pattern can be interpreted by the inset figure, which is the top view of bi-layer structure. Due to the half-period interlayer mismatch, one zirconium atom and four tellurium atoms from adjacent layers form into a unit in TEM structure. Therefore we can measure the lattice constants from TEM image: a = 0.40 nm and c =1.41 nm, which are very close to the results in the literature[19,30].

ZrTe$_5$ bulk crystal was prepared by vapor transport method. Details of crystal synthesis can be found in Supporting Information. After over three weeks of reaction, ZrTe$_5$ crystal was obtained in the form of clusters of long spikes, with the dimensions on the order of 25×1×1mm, as shown in Figure S1(a). XPS results (Figure 1(d)) shows an approximate 1:5 atomic ratio of Zr and Te, with a little trace of transport agent iodine, which may have condensed on the surface of ZrTe$_5$ during temperature cooling down.

Few-layer ZrTe$_5$ was exfoliated by scotch tape and then transferred onto a heavily doped silicon wafer capped with 90nm SiO$_2$. Due to the elongated morphology of bulk ZrTe$_5$ crystal, the exfoliated flakes usually also have slender rectangular shapes. Flake thickness was measured by atomic force microscopy (AFM). Generally speaking, it is difficult to achieve large area of mono-layer ZrTe$_5$ flakes for further study[31]. Yet we can still find some candidate flakes of 3 layers or more with appropriate size for fabrication and characterization. Figure S1 (b) displays an AFM image of an exfoliated ZrTe$_5$ flake with two steps, and the number of layers are identified by AFM to be 2L and 5L respectively.

Angle-resolved Raman spectra were investigated at room temperature. The ZrTe$_5$ crystal structure belongs to the Cmcm ($D_{2h}^{17}$) space group[32], so there are 12×3 vibrational modes at the Γ point of Brillouin zone, among which 18 are Raman-active[33–35]. 2 $B_{2g}$ modes and 6 $A_g$ modes are predicted to be observable when the laser shines perpendicular to (010) surface according to previous study[33,34]. In our experiment, one of the $A_g$ modes is out of the Raman system detector range (39cm$^{-1}$) and another one is relatively weak at room temperature (235cm$^{-1}$), so we focus on the remaining 6 peaks located at 67 cm$^{-1}$, 84 cm$^{-1}$, 113 cm$^{-1}$, 118 cm$^{-1}$, 144 cm$^{-1}$ and 178 cm$^{-1}$. Fresh exfoliated thin ZrTe$_5$ flakes were transferred onto a Si substrate and then mounted onto a rotation stage. Raman spectra were excited by 633nm He-Ne laser which was incident along b-axis and polarized parallel to a-axis. A linear polarizer was placed in front of spectrometer. Only the scattered phonons aligned with polarizer direction can pass through and be collected by the detector. In our experiment, the polarizer is first placed parallel to the incident photon polarization, which is denoted as ac(aa)ac configuration, since this configuration gives the maximum Raman signal. Figure 2(a) presents Raman spectra of ZrTe$_5$ evolving with angles in this configuration. The intensity of six major peaks are extracted and plotted into polar figures in Figure 2(b) – (g). In the meantime, the results of ac(ac)ac configuration (polarizer is perpendicular to the incident light) is also reported in the Supporting Information.

The intensity of different Raman vibration modes can be described as[29]:

$$I = |\mathbf{e}_i \times \mathbf{R} \times \mathbf{e}_s|^2, \qquad (1)$$

where $\mathbf{e}_i$ is the unit vector of incident laser polarization and $\mathbf{e}_s$ stands for the scattering phonon polarization. $\mathbf{R}$ is the Raman tensor for a certain vibration mode. Here we assign the angle between incident laser polarization and a-axis of the lattice of ZrTe$_5$ to be θ. In ac(aa)ac configuration, $\mathbf{e}_i$ is parallel to $\mathbf{e}_s$, thus both of them can be written as: $\mathbf{e}_i = \mathbf{e}_s^T = (\cosθ, 0, \sinθ)$. The Raman tensor can be for $A_g$ and $B_{2g}$ modes are[24]:

$$\boldsymbol{R}_{A_g} = \begin{pmatrix} A & & \\ & B & \\ & & C \end{pmatrix}, \quad \boldsymbol{R}_{B_{2g}} = \begin{pmatrix} & & E \\ & & \\ E & & \end{pmatrix}, \tag{2}$$

Back substitute into eq. (1), for ac(aa)ac configuration, we have:

$$I_{A_g} = (A\cos^2\theta + C\sin^2\theta)^2, \tag{3}$$

$$I_{B_{2g}} = E^2 \sin^2 2\theta \tag{4}$$

The fitted curve based on the calculations are in good agreement with the experimental data. It is straightforward to understand the results for $B_{2g}$ modes (Figure 2(b) - (c)), since both of them have four branches, with maxima and minima at 45° and 90°. In case of $A_g^1$ modes (Figure 2(d)), $A = -C$ gives the period of 90° and maxima at 0°. The rest three modes have only two branches, indicating one of the constants in eq. (3) has to be zero. $A_g^2$ and $A_g^4$ mode plots are fitted with $C = 0$ and $A_g^3$ is fitted with $A = 0$.

In order to investigate the electrical anisotropy of ZrTe₅, we measured angle-resolved DC conductance. Knowing the fact that the growth rate in a-axis is much faster than b- and c-axis, we can preliminarily determine the flake orientation by assigning the long edge of rectangular shape to a-axis. Then we use Raman spectrum to further verify the lattice orientation by the aforementioned method. Prior to fabrication, the flakes were trimmed into circles with the diameter of 10μm by BCl₃/Ar based dry etching. After patterning we can eliminate the non-ideal geometric factors that might be influential on the current flow. 12 contacts were patterned by electron-beam lithography (EBL) and arranged in a circle concentric with the etched flakes in different orientations. The interval between two neighboring contacts are 30°, and 0° is aligned with a-axis. 30 nm Ni and 50 nm Au metal were deposited by electron-beam evaporator. Figure 3(a) schematically illustrates the structure of the device, and the inset of Figure 3(a) is an optical image of the real device. DC conductance was measured between each pair of diagonal contacts as zero back gate bias. A 20% disparity of DC conductance is observed from different angles, as shown in Figure 3(b). Low-field conductivity of an anisotropic material at a certain angle $\theta$ can be decomposed into two orthogonal components[19]: $\sigma_\theta = \sigma_x sin^2\theta + \sigma_y cos^2\theta$, where $\sigma_x$ is the conductivity along (100) direction and $\sigma_y$ is the conductance along (001) direction. Angle-resolved DC conductance measurement reveals the anisotropic transport properties, nevertheless this circular structure underestimates the anisotropic ratio of mobility, which is a fundamental material property, because of fringing current along the conducting path counteracting against anisotropic transport. By designing a more refined L-shaped structure (See Supporting Information), we accurately measured the DC conductance ratio along two primary in-plane axis to be ~1.5.

To acquire more intrinsic understanding of transport properties, we performed low-temperature transport experiments. One of the trademarks of ZrTe$_5$ transport is the anomalous resistivity peak at around 130K (Figure 4(b)). By reducing the temperature, ZrTe$_5$ bulk sample undergoes an anomalous resistivity peak, along with changing sign of Hall coefficient and thermoelectric power[32]. For decades the origin of resistivity peak is under debate and still elusive. Some popular explanation includes formation of charge density waves[36], temperature-induced band movement[11], polaronic behavior[37] and structural phase transition[38]. So far no direct experimental evidence can support any of the proposals. We explored the anomalous resistivity behavior in few-layer ZrTe$_5$ samples. By reducing the flake thickness, the resistivity peak gradually attenuated and eventually disappeared. For thin samples under 10nm, resistivity reduces with temperature monotonically in the full temperature range, as shown in Figure 4(a). Similar results has also been observed in other publications[31]. The Fermi surface of $ZrTe_5$ is complicated, where an electron pocket and a hole pocket contributes to the transport simultaneously. Thus the vanishing of resistivity peak is reckoned as an implication of band structure evolution from bulk to few-layer samples. In bulk sample, Fermi level gradually moves up as temperature drops, so it exhibits a transition from hole-dominant region to electron-dominant region, along with the resistivity peak. Whereas in few-layer sample, the bandgap increased due to reduced flake thickness and electron and hole pockets are no longer entangled together which simplifies the scenario, therefore the R-T curve shows monotonous change.[31]

On the other hand magneto transport experiment is conducted to measure the carrier concentration and Hall mobility along different directions at low-temperature. Standard six-terminal Hall-bar samples (Figure 4(c)) were fabricated with either a-axis or c-axis in longitude direction. The thickness of the devices is ~30nm. Hall mobility and carrier concentrations are obtained by measuring longitudinal resistivity $\rho_{xx}$ and Hall resistivity $\rho_{xy}$ at low temperatures and external magnetic fields. The complexity of Fermi surface makes it difficult to extract mobility and carrier density, since Hall resistivity displays anomalous behavior with both negative and positive slopes (Figure S5(a)). The appropriate methods to extract Hall mobility and carrier density would be to apply two-carrier model. We attempted to extract mobility and carrier density of the dominant carrier type with a simplified method by extracting from the slope of near-linear large B field region (Figure S5(b)). Discuss on correctness of simplified method as well as comparison of these two methods can be found in Supporting Information. Figure 4(d) and (e) show temperature-dependent hole concentration and Hall mobility along two directions. While the carrier concentrations remain the same along two directions, a significant difference arises from Hall mobility. Hole mobility along a-axis is ~3,000cm$^2$/V·s, and along c-axis is ~1,500cm$^2$/V·s along c-axis, which is around a factor of 2 difference.

Particularly worth mentioning is in a 23nm-thick sample electrons become the dominant carrier type and electron Hall mobility reaches over 44,000cm$^2$/V·s along a-axis and strong Shubnikov-de Haas (SdH) oscillations are well developed under 10K as shown in Figure 5(b). The SdH oscillations commence around small B-field around 0.3T (see inset of Figure S5(c)), hence according to observability criteria of SdH oscillations, we can estimate the electron mobility: $\mu_e > 10^4/B \approx 33,000 \; cm^2/V \cdot s$, which is a reasonable approximation to the value we extracted from Rxy. The inconsistency of Hall resistivity behavior in different samples was presumed to arise from the thickness-induced bandgap widening and two types of carriers are disentangled at Fermi surface. More details are discussed in Supporting Information. Hall mobility is extracted from Hall resistivity curve and plotted in Figure 5(a). By subtracting a smooth background, SdH oscillation amplitude $\Delta R_{xx}$ (Figure 5(b)) is obtained. Oscillation amplitude versus temperature at different B-fields (Figure 5(c)) are fitted by Lifshitz-Kosevich equation[39]:

$$\Delta R_{xx} \propto \frac{2\pi^2 k_B m^* T/\hbar eB}{\sinh(2\pi^2 k_B m^* T/\hbar eB)}, \qquad (6)$$

from which we can extract effective mass to be ~0.032m$_0$ (Figure 5(d)). The extracted effective mass is 0.032m$_e$ which is very close to the results in other works[10,16,17,19].

In summary, we synthesized bulk ZrTe$_5$ single crystal by vapor transport method and mechanically peeled it down to few-layer flakes. We applied XPS to confirm the elementary composition and TEM image to elucidate the lattice structure. Raman spectroscopy reveals different phonon dispersion modes at the Γ points and angle-resolved peak intensity behavior is in good agreement with calculations. We also measured DC conductance along different axis with a significant anisotropic ratio of around 1.5 at room temperature. The anomalous resistivity peak at low temperature vanished in few-layer samples, which implies the disentanglement of hole pocket and electron pocket at Fermi surface resulting from bandgap widening as thickness reduces. Finally we carried out Hall measurement at low temperatures to provide an insightful perspective of anisotropic transport. Hall mobility along a- and c-axis are 3,000 and 1,500 cm$^2$/V·s, respectively. SdH oscillations were observed in the sample where electron mobility reaches up to 44,000 cm$^2$/V·s. The strong optical and electrical anisotropy not only gives us a detailed understanding of this material system, but also offers a solid method to determine crystal orientation.

**Supporting information**

Details of crystal growth, few-layer AFM image, additional optical and electrical results, device fabrication, discussion of Hall resistivity and mobility extraction and two carrier model can be found in supporting information.

**Acknowledgements**

One of the authors (P.D.Y.) would like to thank W. Pan and Z. Jiang for the valuable discussion, which initiates this research work. This material is based upon work partly supported by AFOSR/NSF EFRI 2-DARE Grant No. 1433459-EFMA and partly supported by SRC GRC program. The low temperature measurements were performed at the National High Magnetic Field Laboratory (NHMFL), which is supported by National Science Foundation Cooperative Agreement No. DMR-1157490, the State of Florida, and the US Department of Energy. The authors would like to thank T Murphy, J-H Park, and G Jones for experimental assistance.

**Figures**

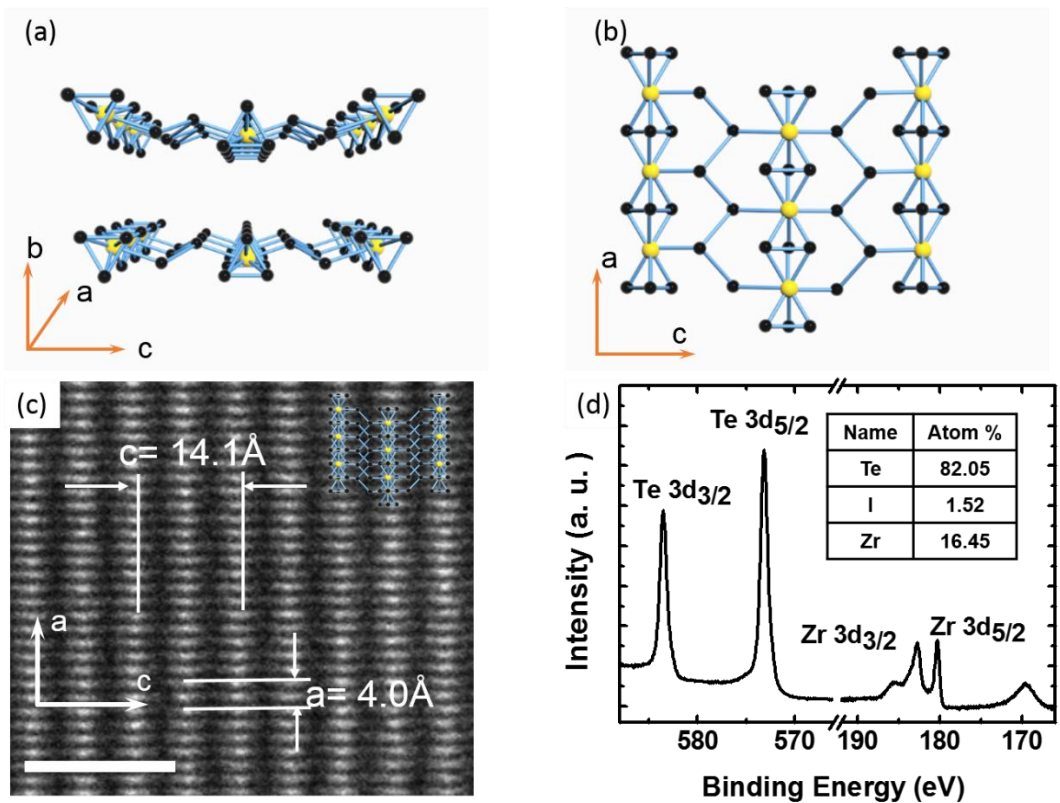

**Figure 1**. Crystal structure of few-layer $ZrTe_5$. (a) Perspective side view of few-layer $ZrTe_5$, (b) top view of mono-layer $ZrTe_5$. (c) Transmission electron microscopy (TEM) image of $ZrTe_5$ (010) surface. The scale bar is 2nm. Inset: schematic structure of bi-layer $ZrTe_5$ from (010) surface, which matches the TEM lattice period. (d) X-ray photoemission spectroscopy (XPS) results of binding energies of Zr and Te. Inset table summarizes the atomic ratio of measured sample.

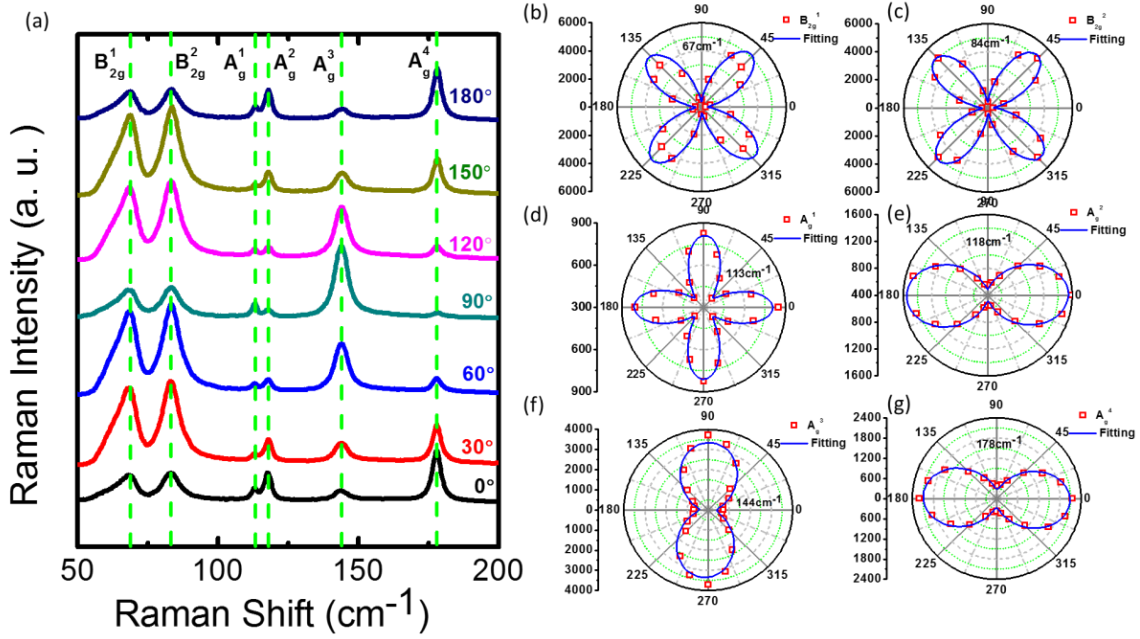

**Figure 2** (a) Raman spectra of a fresh exfoliated flake evolving with rotating angle between the flake and incident polarization. (b) – (g) are corresponding to $B_{2g}^1$, $B_{2g}^2$, $A_g^1$, $A_g^2$, $A_g^3$ and $A_g^4$ Raman modes which are located at 67 cm$^{-1}$, 84 cm$^{-1}$, 113 cm$^{-1}$, 118 cm$^{-1}$, 144 cm$^{-1}$ and 178 cm$^{-1}$.. The fitted curves are described as equation (3) and (4).

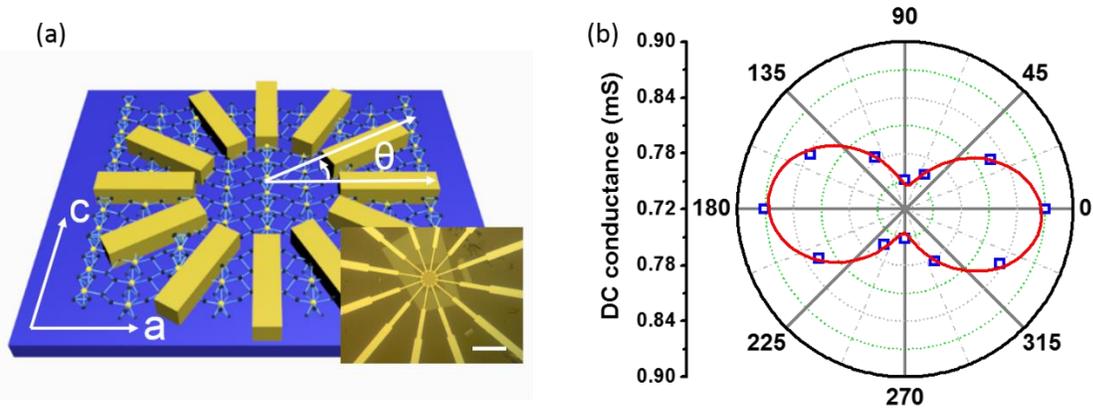

**Figure 3.** Angle-resolved DC conductance measurements on ZrTe5 flake. (a) Schematic view of device structure. Inset: optical image of a real device. Scale bar is 10μm. (b) Angle-dependent DC conductance. The data points are fitted with the equation: $\sigma_\theta = \sigma_x sin^2\theta + \sigma_y cos^2\theta$.

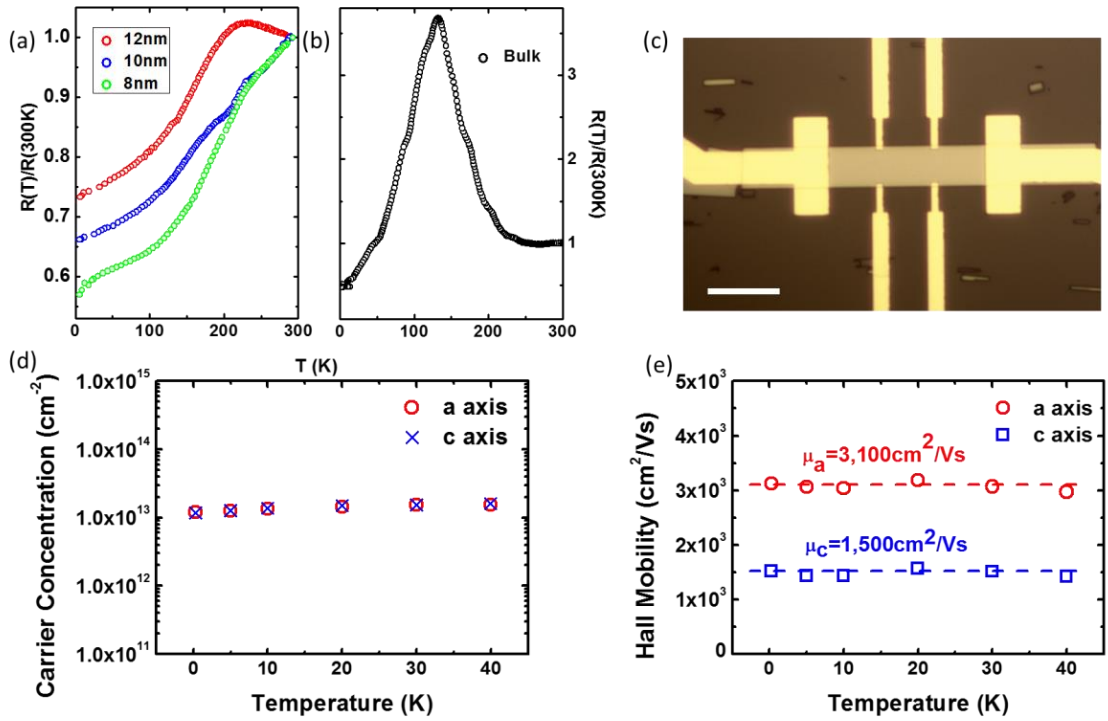

**Figure 4.** Low-temperature transport measurements. Resistivity vs. temperature curve on (a) few-layer and (b) bulk $ZrTe_5$ samples. The resistivity is normalized based on room temperature values. (c) Optical image of the Hall-bar device. The scale bar is 10 μm. Temperature dependence of (d) carrier concentration and (e) Hall mobility along two principal axis.

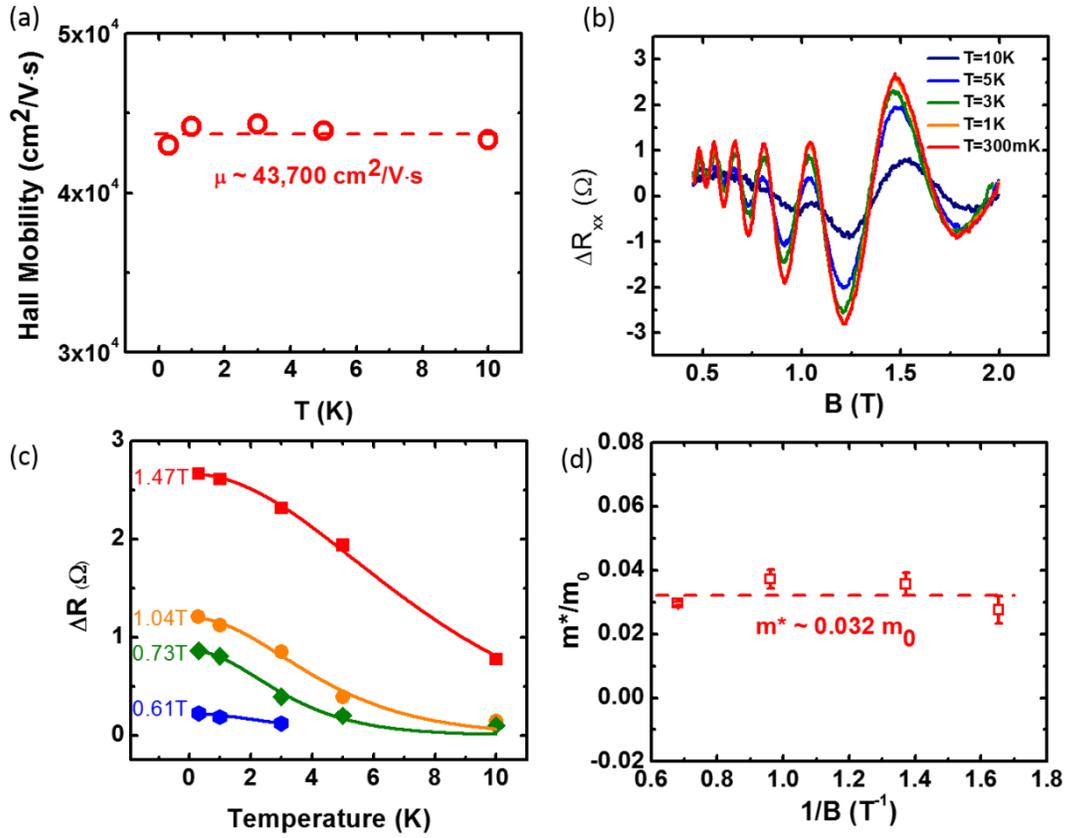

**Figure 5** (a) Hall mobility extracted from the slope of Rxy. (b) SdH quantum oscillations as a function of B field after subtracting from a smooth background. (c) Temperature dependence of oscillation amplitudes. (d) Effective mass versus 1/B extracted from (c) using eq. (6).

TOC only

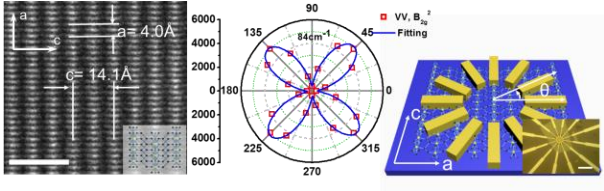

Supporting Information

# Observation of Optical and Electrical In-plane Anisotropy in High-mobility Few-layer ZrTe$_5$

Gang Qiu[1,4], Yuchen Du[1,4], Adam Charnas[1,4], Hong Zhou[1,4], Shengyu Jin[2,4], Zhe Luo[3,4], Dmitry Y. Zemlyanov[4], Xianfan Xu[3,4], Gary J. Cheng[2,4], Peide D. Ye[*,1,4]

[1] School of Electrical and Computer Engineering, Purdue University, United States

[2] School of Industrial Engineering, Purdue University, United States

[3] School of Mechanical Engineering, Purdue University, United States

[4] Birck Nanotechnology Center, Purdue University, United States

\* Corresponding author

E-mail: yep@purdue.edu.    Fax: 765-496-7443

**Crystal growth**

The ZrTe$_5$ single crystal is synthesized by vapor transport method[1,2]. Stoichiometric amount of Zirconium (foil, 99.98%, Sigma-Aldrich), Tellurium (99.9999%, Sigma-Aldrich), and 30mg Iodine (99.999%, Sigma-Aldrich) were loaded into a quartz ampoule with wall thickness 3mm and inner diameter 6mm. The ampoule was evacuated to under $10^{-4}$ Torr and then sealed with propane torch. A two-zone furnace was set at 480 °C and 400 °C for two area separately and the starting material was located at the hot end. After 500 hours of reaction time, large bulk ZrTe$_5$ is crystallized as shown in Fig. S1(a).

**Additional angle-resolved Raman results**

ac(ac)ac configuration refers to the Raman experiment setup where the polarization of incident laser phonon is perpendicular to that of the scattered phonon. Fig. S2(a) shows Raman spectra of few-layer ZrTe$_5$ evolving with angle in ac(ac)ac configuration. Unlike in ac(aa)ac configuration where 6 Raman modes are prominent, in ac(ac)ac configuration three $A_g$ mode signals are too weak peak fitting. So we only plot polar figures of $B_{2g}^1$, $B_{2g}^2$ and $A_g^1$, as shown in Fig. S2(b) – (d). In ac(ac)ac configuration, $\boldsymbol{e}_i = (\cos\theta, 0, \sin\theta)$, $\boldsymbol{e}_s^T = (-\sin\theta, 0, \cos\theta)$, and Raman tensor is the same as in eq. (2). Back substitute into eq. (1) we have:

$$I_{A_g} = \frac{(C-A)^2}{4}\sin^2 2\theta, \quad (1)$$

$$I_{B_{2g}} = E^2\cos^2 2\theta \quad (2)$$

These equations indicates both $A_g$ and $B_{2g}$ modes have four branches; $B_{2g}$ has maxima at 0° and minima at 45°, whereas $A_g$ modes is the other way around: exactly matches the experiment results.

**Device fabrication**

Here we introduce fabrication process of L-shaped device as an example. Flakes were exfoliated by scotch tape and transferred onto 90nm SiO$_2$/Si substrates as shown in Fig. S3(a). In order to eliminate geometric non-ideality in anisotropic study, we trimmed the device into desired shapes for symmetry before fabrication. Electron-beam lithography was applied to open a window on the flake except the L-shaped strip region. Then the flake was dry etched by Ar/BCl$_3$ for 1 minute (Fig. S3(b)). Again we use e-beam lithography to pattern a shared source and two drain contacts, followed by electron beam evaporation of 40nm Ni and 40nm Au as metal contacts(Fig. S3(c)).

**Additional electrical anisotropic results**

In order to measure the conductivity and mobility ratio accurately, a second type of L-shaped devices were fabricated. Flakes were etched into 2μm wide L-shaped strips, with two edges along a- and c-axis. Two drain and one shared source metal contacts were deposited as shown in Fig. S4(a). By this design we have a pair of field-effect transistors with channel orientation along a- and c-axis on the same flake. For simplicity, these two types of device are denoted as device A and device C, based on the channel orientation, in the rest of the paper. Fig. S4(b) shows the transfer curve of two devices fabricated on the same L-shaped strip with the flake thickness of 7.8nm. Note the fact that as a three dimensional Dirac semimetal, even the monolayer of ZrTe$_5$ is predicted to have a small bandgap of 100meV [3]. So it is reasonable that the FET device only has gate modulation of around 30% at room temperature. Maximum current in device A is 1.5 times larger than device C, which agrees well with our previous prediction: the anisotropic conductivity ratio is enhanced by eliminating fringe current in L-shaped devices. Figure S4(c) shows the output curve of device A and device C. The linear curve indicates that the contact is ohmic type and thus we can ignore the small contact resistance and estimate sheet resistance directly from Id-Vd curve: $R_{sh\_a} = 1.40 k\Omega/\square$, $R_{sh\_c} = 2.09 k\Omega/\square$. The DC conductance along a-axis is 50% percent larger than that along c-axis.

**Hall resistivity anomaly and mobility extraction**

The complicated Fermi surface of ZrTe$_5$ makes it difficult to interpret Hall resistivity curve and extract mobility and carrier density. We observed anomalous Hall resistivity behavior in near zero B-field region in the sample that we used to extract anisotropic transport behavior. The Hall resistivity curve shows both positive and negative slope (Figure S5(a)), which indicates that this anomaly arises from nesting of hole pocket and electron pocket.

The appropriate way to extract mobility and carrier concentration would be to apply the classical two-carrier model. Classical magneto-transport in a two-carrier semimetal is given by[4, 5]:

$$\rho_{xx} = \frac{(n\mu_e + p\mu_h) + (n\mu_e\mu_h^2 + p\mu_e^2\mu_h)B^2}{e[(n\mu_e + p\mu_h)^2 + (p-n)^2\mu_h^2\mu_e^2 B^2]} \quad (3)$$

$$\rho_{xy} = \frac{(p\mu_h^2 - n\mu_e^2)B + (p-n)\mu_h^2\mu_e^2 B^3}{e[(n\mu_e + p\mu_h)^2 + (p-n)^2\mu_h^2\mu_e^2 B^2]} \qquad (4)$$

For instance, Figure S5(a) shows a sample (45nm) with a typical non-linear B field dependence of R_xy. By fitting the curve with equations above, we can extract the mobility and carrier density of two carriers: $p = 2.0 \times 10^{13} cm^{-2}$, $\mu_h = 1,100 cm^2/V \cdot s$, $n = 8.8 \times 10^{11} cm^{-2}$, $\mu_e = 53,000 cm^2/V \cdot s$. This result draws a detailed band picture: a high-density low-mobility hole pocket and a low-density high-mobility electron pocket simultaneously contribute to the transport (red line in Figure S5(d)).

In our manuscript we used a simplified method by extracting the mobility of the dominant carrier type from the slope at large-B field where it is almost linear (Figure S5(b)). By this method we acquired the hole mobility and carrier density to be: $p = 2.1 \times 10^{13} cm^{-2}$, $\mu_h = 1,600 cm^2/V \cdot s$, which is a good approximation of the previous two-carrier model within a reasonable range. Actually if we compare these two methods in principle, when B is large enough we can omit the first term in both denominator and numerator in equation (3) and (4), and therefore the equations can be rewrite to classical Hall resistivity expression. This further verifies that it is sound to extract Hall slope at large B fields.

The Hall resistivity curve of a high mobility sample is shown in Figure S5(c). In this sample the Hall resistivity does not demonstrate anomalous two-carrier behavior but instead shows a simple negative slope from 0 to large B field, which means only electron transport dominates in this sample. A tentative explanation is that in this thin sample (23nm), the reduced thickness increases the bandgap and disentangles the hole and electron pockets, so only the electron branch is at Fermi surface. In this case we simply used near-zero B field region where is almost consistently linear to extract the Hall mobility. Noted that the high electron mobility extracted in this sample, which is an intrinsic property of this material, is similar to that in previous sample extracted by two carrier model, despite the fact that the $R_{xy}$ curves are tremendously different in these two samples. The band diagram in Figure S5(d) explains the paradox why different samples of the same material behave distinctively in magneto-transport – in a thick sample the mix of two carriers results in anomalous Hall resistivity behavior whereas in thin samples only electron transport is dominant.

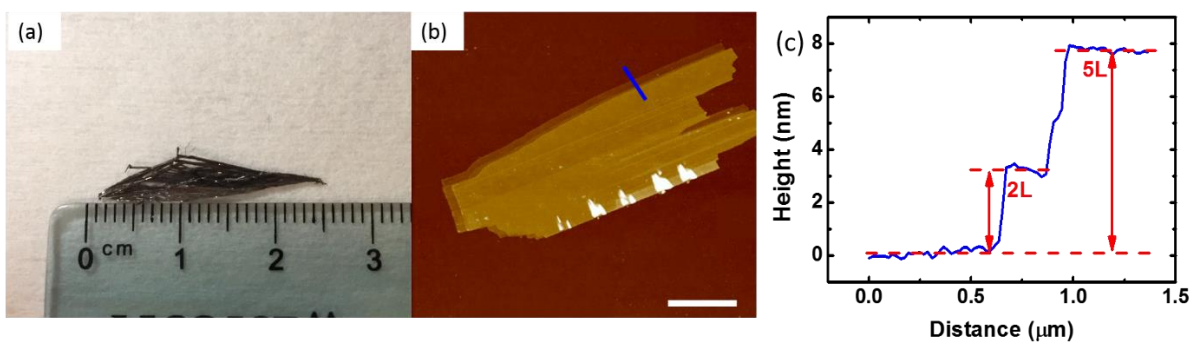

**Figure S1.** (a) As-grown ZrTe$_5$ single crystal with the dimensions of 25×1×1mm. (b) atomic force microscopic (AFM) images of a mechanically exfoliated few-layer flake. The scale bar is 2μm. The number of two regions is 2L and 5L respectively. (c) Height profile of the exfoliated flake measured along the blue line in (b).

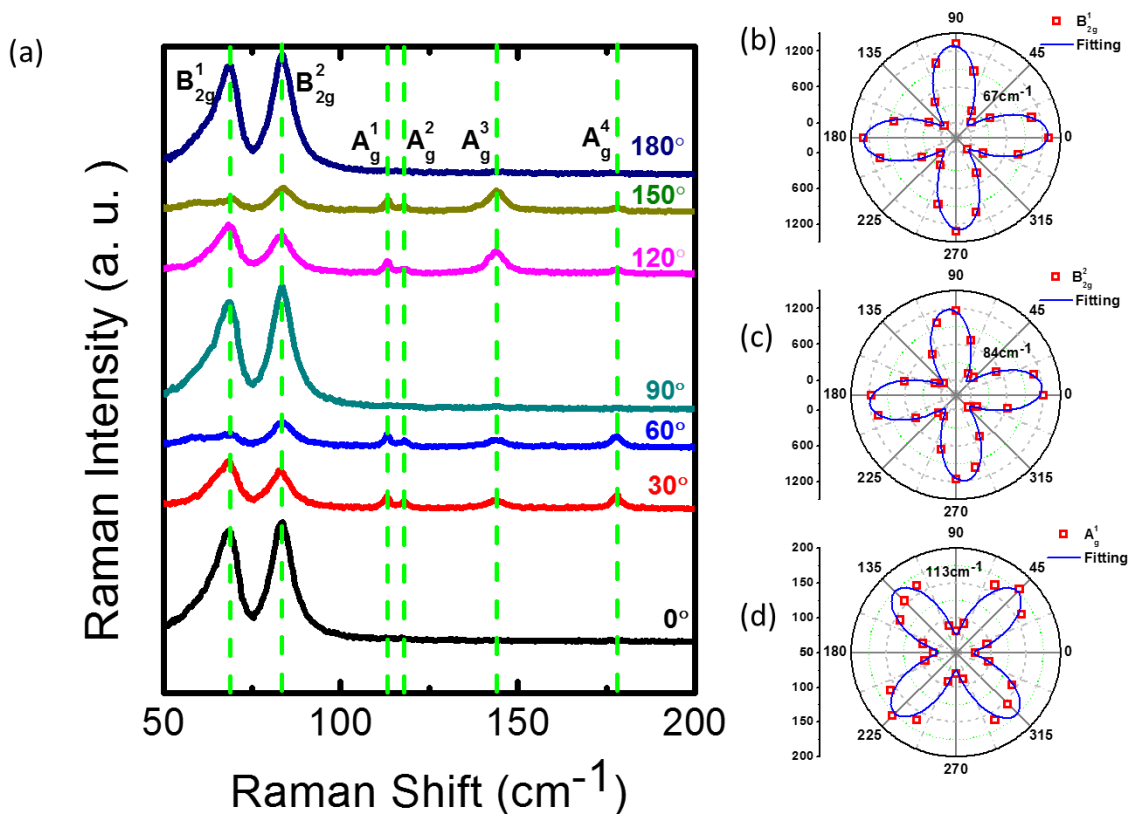

**Figure S2.** (a) Angle-resolved Raman spectra in ac(ac)ac configuration. (b) – (d) are corresponding to $B_{2g}^1$, $B_{2g}^2$ and $A_g^1$ Raman modes. The rest three modes are too weak for peak fitting.

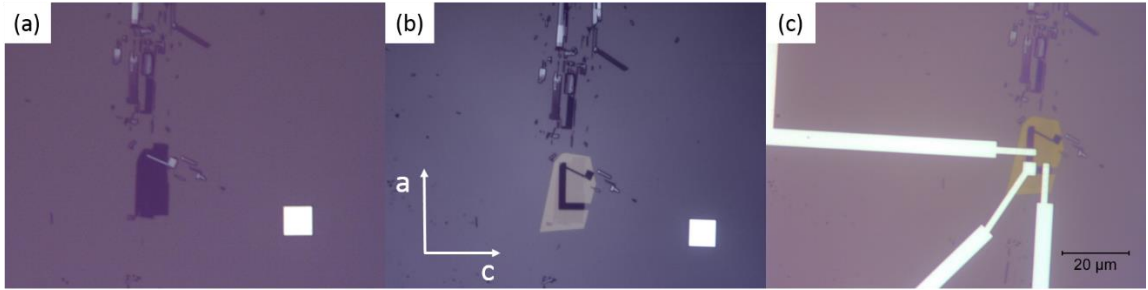

**Figure S3.** L-shaped device fabrication process. (a) As-exfoliated flakes on 90nm SiO$_2$/Si substrates. (b) electron-beam lithography is applied to open the window over the flake except L-shaped region, followed by Ar/BCl$_3$ dry etching. (c) 40nm Ni and 40nm Au is deposited as metal contacts.

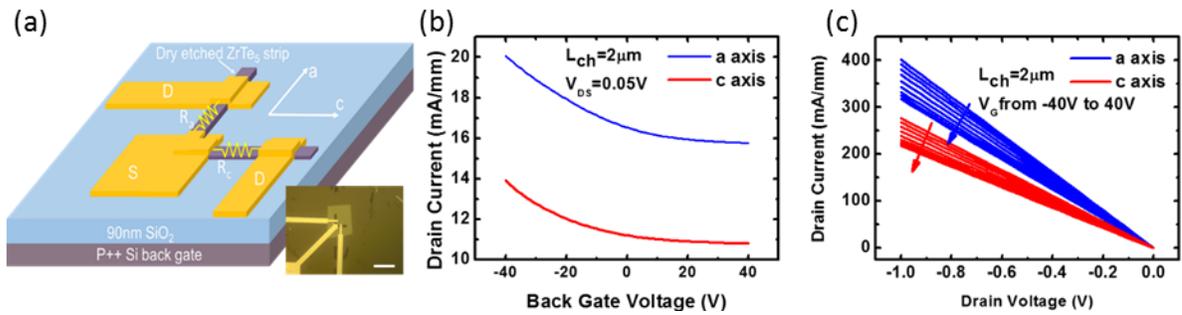

**Figure S4.** (a) L-shaped device for measuring electrical anisotropic properties. Inset: Optical image of an 8-nm-thick device. The scale bar is 10 μm. (b) Transfer and (c) output curves of two perpendicular devices with channel length of 2μm.

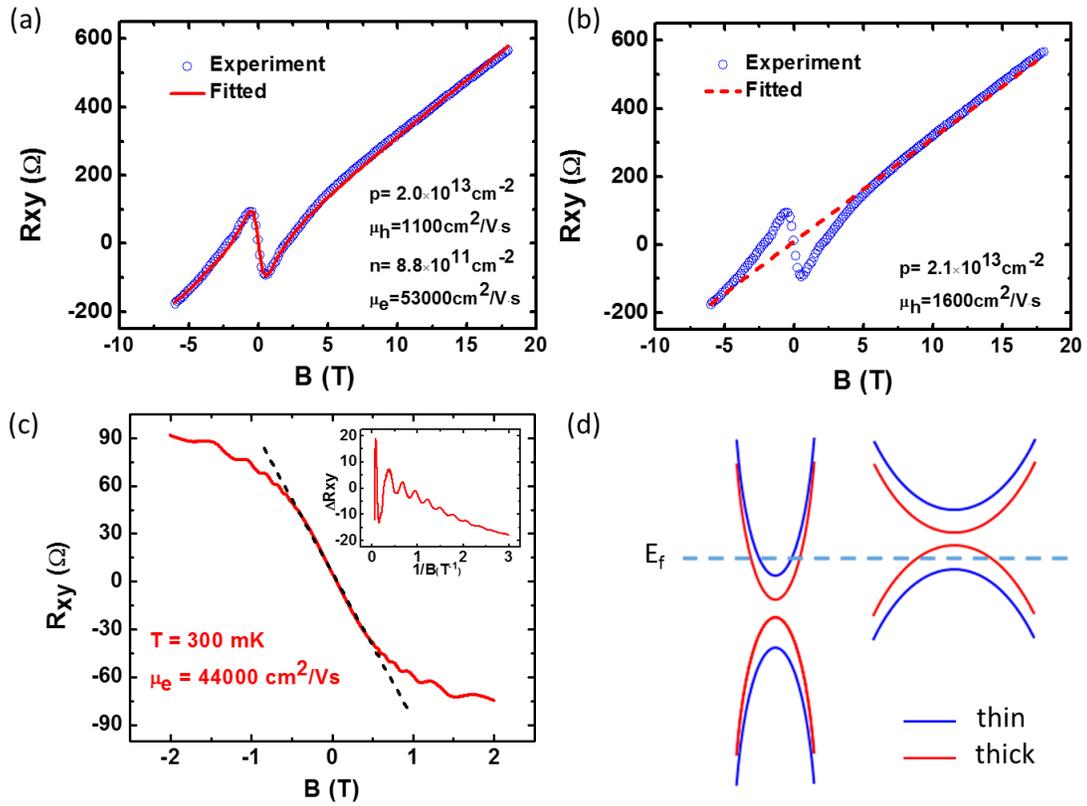

**Figure S5.** (a) Hall resistivity anomaly and two-carrier model fitting; (b) Large B-field mobility extraction; (c) Hall resistivity curve of a high mobility sample (23nm). Mobility is extracted from near zero B-field region where it is almost linear. Inset: The quantum oscillation with a single 1/B frequency is to further confirm a single dominant carrier of high mobility electron. (d) Schematic of the band structure of $ZrTe_5$ with the two carrier model. Red line and blue line represent thick and thin samples respectively. The Dirac characteristics in band-structures are ignored here.